\documentclass[12pt,preprint]{aastex}

\slugcomment{\today}

\def\ea{{\it et al.\,}}
\def\be{\begin{equation}}
\def\ee{\end{equation}}
\def\eg{{\it e.g.,\,}}

\begin{document}
\title{Triple Experiment Spectrum of the Sunyaev-Zeldovich Effect in
the Coma Cluster: $H_0$}

\author{ E.S. Battistelli \altaffilmark{1}, M. De Petris,
L. Lamagna, G. Luzzi, R. Maoli, A. Melchiorri, F. Melchiorri, A.
Orlando, E. Palladino, G. Savini} \affil{ Department of Physics,
University "La Sapienza", P.le A. Moro 2, 00185, Rome, Italy}

\author{Y. Rephaeli, M. Shimon}
\affil{School of Physics and Astronomy, Tel Aviv University, Tel
Aviv, Israel}

\author{M. Signore}
\affil{LERMA, Observatoire de Paris, Paris, France}

\author{S. Colafrancesco}
\affil{INAF - Osservatorio Astronomico di Roma, Via Frascati 33,
00040, Monteporzio, Italy}

\altaffiltext{1}{elia.stefano.battistelli@roma1.infn.it}

\begin{abstract}

The Sunyaev-Zeldovich (SZ) effect was previously measured in the
Coma cluster by the Owens Valley Radio Observatory and Millimeter
and IR Testa Grigia Observatory experiments and recently also with
the {\it Wilkinson Microwave Anisotropy Probe} satellite. We
assess the consistency of these results and their implications on
the feasibility of high-frequency SZ work with ground-based
telescopes. The unique data set from the combined measurements at
six frequency bands is jointly analyzed, resulting in a best-fit
value for the Thomson optical depth at the cluster center,
$\tau_{0}=(5.35 \pm 0.67)\times 10^{-3}$. The combined X-ray and
SZ determined properties of the gas are used to determine the
Hubble constant. For isothermal gas with a $\beta$ density profile
we derive $H_0 = 84 \pm 26\,\, km/(s\cdot Mpc)$; the ($1\sigma$)
error includes only observational SZ and X-ray uncertainties.
\end{abstract}


\section{Introduction}

The Sunyaev-Zeldovich (SZ) effect \cite{SunZel72} constitutes a
unique and powerful cosmological tool (for reviews, see Rephaeli
1995a, Birkinshaw 1999, Carlstrom, Holder, \& Reese 2002). Many
tens of high quality images of the effect have already been
obtained with interferometric arrays operating at low frequencies
on the Rayleigh-Jeans side of the spectrum. Multifrequency SZ
measurements were made of only very few clusters, so the potential
power of spectral diagnostics has not yet been sufficiently
exploited to reduce signal confusion and other errors. Systematic
uncertainties are the main hindrance to the use of the effect as a
precise cosmological probe. These can be most optimally reduced
when high spatial resolution X-ray and spectral SZ measurements of
a large sample of {\it nearby} clusters are available in the near
future. The multifrequency capability of many upcoming
ground-based bolometric array projects is crucially important for
the separation of the SZ signal from the large confusing
atmospheric signals.

The SZ effect was measured in the Coma Cluster with the
ground-based Owens Valley Radio Observatory (OVRO; Herbig,
Lawrence, \& Readhead 1995) and Millimeter and IR Testa Grigia
Observatory (MITO; De Petris \ea 2002) telescopes. The {\it
Wilkinson Microwave Anisotropy Probe} ({\it WMAP}) team has
recently reported measurement of the effect in Coma at 61 and 94
GHz \cite{ben03b}; this is the first measurement of the SZ effect
from a satellite. Together, these measurements yield the first SZ
spectrum with six spectral bands. The relatively wide spectral
coverage obtained with ground- and space-based telescopes allows a
much-needed gauge of the impact of atmospheric emission on
high-frequency ground-based observations of the SZ effect. The
quality of the spectral results allows a meaningful determination
of the Hubble constant, $H_0$. Even though this important
parameter is thought to have been determined quite precisely from
the {\it WMAP} first-year sky survey (Bennett \ea 2003a), it is,
of course, very much of interest to measure it independently by
various methods. The determination of $H_0$ from cluster SZ and
X-ray measurements yields additional insight beyond what is
offered by an all-sky cosmic microwave background (CMB) anisotropy
survey: in principle, the cluster SZ \& X-ray (SZ-X) method
provides a test of both the isotropy of the cosmological expansion
and its temporal character. In this Letter we briefly discuss the
qualitative deductions that can be drawn from the general
agreement between the ground-based and {\it WMAP} results, and use
the full database on Coma to determine $H_0$.

\section{SZ Measurements of the Coma Cluster}

\begin{figure}
\plotone{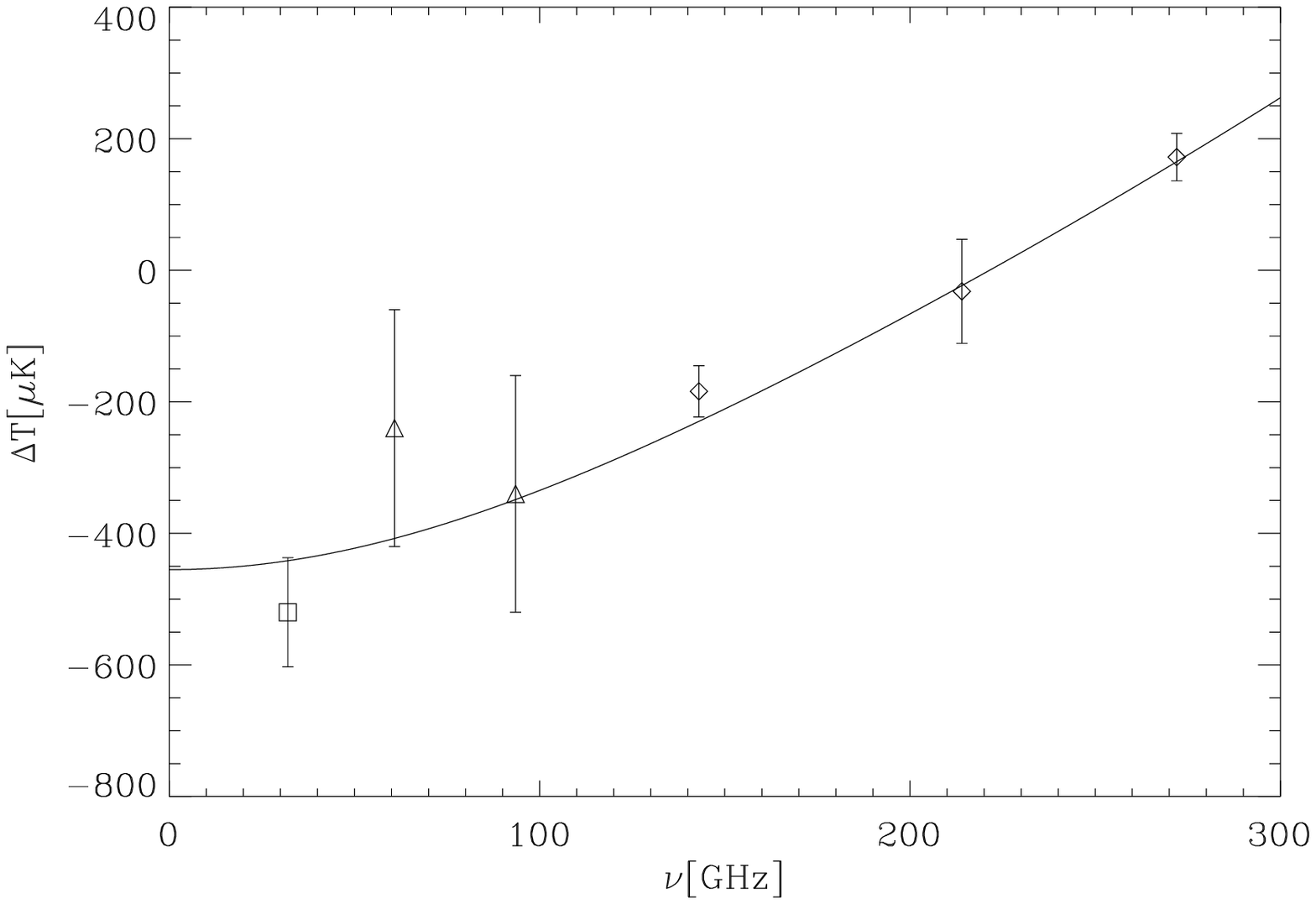} \caption{SZ spectrum of the Coma cluster. { \it
Solid line}: best fit spectrum (assuming isothermal gas with $kT =
8.21$ keV) to the combined MITO (diamonds), OVRO (square), and
{\it WMAP} (triangles) measurements, corresponding to $\tau =
(5.35 \pm 0.67) \times 10^{-3}$.\label{fig_1}}
 \label{fig:sz} \end{figure}

The rich nearby Coma Cluster has long been a prime target of SZ
observations (\eg Birkinshaw, Gull, \& Northover 1981, Silverberg
1997); the effect was detected towards Coma with the 5.5 m OVRO
telescope, with a 7'.3 beam, operating at 32 GHz. A long series of
drift scans with the MITO 2.6 m telescope, which had a 16' beam,
led to the detection of the effect at three spectral bands
centered on 143, 214, and 272 GHz. Even though the {\it WMAP}
telescope is not optimized for SZ observations, the effect was
detected with the W- and V-band radiometers at 61 and 94 GHz, with
beam sizes of 20' and 13', respectively \cite{ben03b}. The
predicted SZ size of Coma is more than $\sim 30'$, so these
measurements were not substantially affected by beam dilution.
Results of the measurements are listed in Table \ref{tab:data};
these reflect a slight revision of our previously reported SZ
signals as result of an improved calibration of the the MITO
measurements \cite{sav03}. The SZ spectrum of Coma is shown in
figure \ref{fig:sz}.

Note that the MITO error bars are about a factor 3 smaller than
those of the {\it WMAP} data; this is mainly due to the difference
in observing time. Given the low signal-to-noise ratios (S/Ns) of
the two {\it WMAP} data, the main advantage of these satellite
observations lies in the capability of relatively precise
calibration through measurements of the modulation of the CMB
dipole anisotropy. This is estimated to be $\sim 0.5\%$ by the
{\it WMAP} team, as compared with $\sim 10\%$ for MITO and from
$\sim 3\%$ to $\sim 10\%$ for OVRO. It is interesting that the
relative calibration of MITO and {\it WMAP} are in agreement to
within $\sim 10\%$. This result provides further support for
ground-based SZ observations, which are susceptible of substantial
uncertainties because of atmospheric emission and use of planetary
emission for signal calibration. However, because of the poor S/N
of the {\it WMAP} data, the situation with regard to this
calibration method is still far from being satisfactory. We also
note that an attempt to explain the (small) discrepancy between
the MITO and OVRO measurements as due to confusing CMB anisotropy
signal (De Petris {\it et al.} 2002) seems less likely in light of
the {\it WMAP} results which imply that such a signal is
relatively insignificant.

In the relativistically accurate calculation of $\Delta I$,
the intensity change in Compton scattering, its dependence on
the gas temperature is no longer linear (Rephaeli 1995b), and
can be written in the form (Battistelli \ea 2002)
\be
\Delta I = { 2 k^{3} T^{3} \sigma_{T} \over h^{2}c^{2} } {x^4e^x
\over (e^x -1)^2} \int n_{e}
\biggr[\frac{kT_{e}}{mc^2}f_{1}(x)-\frac{v_{r}}{c} + R(x, T_{e},
v_{r}) \biggl ] dl \, , \ee where $T$ is the CMB temperature,
$\sigma_{T}$ is the Thomson cross section, $x = h\nu /kT$ is the
nondimensional frequency, $n_{e}$ and $T_{e}$ are the electron
density and temperature, $v_{r}$ is the line of sight (LOS)
component of the cluster (peculiar) velocity in the CMB frame, and
$f_{1}(x) = x(e^x +1)/(e^x -1) - 4$. The integral is over the LOS
through the cluster. Both the thermal and kinematic components of
the effect are included in equation (1), separately in the first
two terms, and jointly in the function $R(x,T_{e},v_{r})$.
Analytic approximations to the results of the relativistic
calculations have been derived to various degrees of accuracy by
expansion in powers of $kT_{e}/mc^2$ and $v_{r}/c$ (\eg,, Itoh \ea
1998; Nozawa, Itoh, \& Kohyama 1998; Shimon \& Rephaeli 2003;
Colafrancesco, Marchegiani, \& Palladino 2003). If the gas is
isothermal and the cluster velocity is sufficiently small so that
the velocity-dependent terms are negligible (at the relevant
frequencies), then $\Delta I \propto \tau$, and a one parameter
fit can be performed to the spectral SZ measurements in order to
determine the value of $\tau$. Since the cluster velocity is
unknown, the contribution of the kinematic terms can then be
treated as a source of systematic error.

The Coma cluster was observed by all X-ray experiments; we use
values of the gas parameters as deduced by Mohr, Mathiesen, \&
Evrard (1999) from {\it ROSAT} observations. The emission-weighted
gas temperature is $kT_e = 8.21 \pm 0.16$, and the gas core
radius, index of the $\beta$-model density profile, $n_{e}(\theta)
= n_{e0}(1 + \left(\theta/\theta_{c}\right)^{2})^{-3\beta/2}$
(where $\theta$ is the angular radial variable), and central
surface brightness are $\theta_{c} = 9'.97 \pm 0'.67$, $\beta =
0.705 \pm 0.046$, and $S_{X,0}=(4.65 \pm 0.14)\times 10^{-13} \,
erg\; s^{-1}\; cm^{-2}\;arcmin^{-2}$, respectively. These values
are consistent with the results from more recent {\it XMM}
measurements.

Best fitting the six data points in Figure \ref{fig:sz} by the
calculated spectrum (taking $kT_e = 8.21 \pm 0.16$) we obtain the
central value of the Thomson optical depth, $\tau_{0}=(5.35\pm
0.67) \times 10^{-3}$, where the error reflects a 68\% confidence
level uncertainty associated with the $\chi^{2}$ minimization
procedure and the uncertainty due to the error in the observed
electron temperature which contributes $\sim 0.10$.

\section{The Hubble constant, $H_0$}

As is well known, the combination of SZ and X-ray measurements
yields a determination of the cluster angular diameter distance,
$d_A$, from which $H_0$ can then be deduced by a comparison with
the theoretical expression for $d_A$. The equation used to
determine $d_A$ is obtained by combining the formulae for $\Delta
I$ and the X-ray surface brightness, $S_{X}$, through the
elimination of the dependence of these quantities on the central
electron density. This procedure has been discussed in detail in
numerous works (\eg Holzapfel {\it et al.} 1997, Hughes \&
Birkinshaw 1998, Furuzawa {\it et al.} 1998, Reese {\it et al.}
2000, Mason, Myers, \& Readhead 2001) and has by now become
standard. Here we employ this method to determine $H_0$ for the
first time from multispectral measurements of the Coma cluster.

Assuming an isothermal gas with a $\beta$-model density profile
(Cavaliere \& Fusco-Femiano 1976), the intensity change is $\Delta
I \propto \tau_{0} (1+ (\theta / \theta_{c})^{2})^{1/2-3/2\beta}$,
and its explicit dependence on $x$ and $T_{e}$ is determined by
the form of the analytic approximation to the exact relativistic
calculation. The X-ray surface brightness is
\be
S_{X}= \frac{d_{A}}{4\pi(1+z)^{4}}\int \sum_{j}n_{e} n_{j}
Z_{j}^{2} \Lambda _{Bj} d \zeta = S_{X,0} [1+ (\theta / \theta
_{c})^{2}]^{(1/2-3\beta)}, \label{eq:SX} \ee where $z$ is the
cluster redshift, and $\zeta=l/d _{A}$ is the cluster path length
along the LOS in terms of the angular diameter distance. The sum
is over the various ionic species (j), with densities $n_{j}$ and
charges $Z_{j}$. The thermal bremmstrahlung emissivity in units of
$erg~s^{-1}~cm^{3}$ is
\be
\Lambda_{Bj}= 1.426 \times 10^{-27} T_{e}^{1/2}
\int_{\epsilon_{1}}^{\epsilon_{2}} e^{-\epsilon}
G_{Z_{j}}(T_{e},\epsilon) d\epsilon , \label{eq:la} \ee where
$\epsilon=E/kT_{e}$, $E$ is the emitted X-ray photon energy, and
$G_{Z _{j} }$ is the Gaunt factor. We have used the accurate
analytic formula for the Gaunt factor given by Itoh {\it et al.}
(2000, 2002).

A comparison of the deduced value of $d_A$ with the theoretical
expression in a cosmological model with a cosmological constant
then yields the value of $H_0$ when the values of the cosmological
density parameters $\Omega_{M}$ and $\Omega_{\Lambda}$ are
specified. The final expression for $H_{0}$ is
\be
H_{0}= 4 \pi c (1+z)^{4} \frac{\sigma _{T}^{2}S_{X,0} \theta_{c}}
{\Lambda_{B}} \frac{(\int F_{n})^{2}}{\tau^2 \int F_{n}^{2}}
g(z,\Omega_{M},\Omega_{\Lambda}), \label{eq:H} \ee where the
integrals over the LOS have been performed up to a maximal radius
of $10(\theta_{c})$, by defining $F_{n}=(1+\xi^{2})^{-3\beta/2}$
where $\xi=\zeta/\theta_{c}$. The function
$g(z,\Omega_{M},\Omega_{\Lambda})$ is defined as
\be
g(z,\Omega_{M},\Omega_{\Lambda}) = \int_{0}^{z}\frac{dz'}
{\sqrt{\Omega_{M}(1+z')^{3}+\Omega_{\Lambda}}}. \ee The values
$\Omega_{M}=0.27$ and $\Omega_{\Lambda}=0.73$ have been adopted in
the computation. Doing so we obtain
\be
H_{0} = 84 \pm 26 \,\, km/(s\, Mpc), \ee where the error is
determined combining the observed SZ ($1\sigma$) uncertainty and
the uncertainties in the X-ray data. These errors contribute
comparably to the overall observational uncertainty in the deduced
value of $H_{0}$.

To account for the possibility of a nonisothermal profile, we
consider a polytropic gas model such that the temperature spatial
distribution is
$T_{e}(r)=T_{e0}[1+(r/r_{c})^{2}]^{-3\beta(\gamma-1)/2}$, with the
('polytropic') index $\gamma$ considered as a free parameter to be
determined (mostly) from X-ray measurements. Ideally, when
spatially resolved measurements are available of the X-ray
spectrum and surface brightness profile, it is possible to
determine values of the central temperature, central density, core
radius, and the indices of the $\beta$ and polytropic profiles by
direct fits to the data. This complete procedure will be feasible
when the full {\it XMM} data set on Coma becomes available. For
now, we use the previously determined {\it ROSAT} parameters to
estimate the implied change in the value of $H_{0}$ for an {\it
assumed} range of values of $\gamma$. For a given value of
$\gamma$ we determine the central temperature by keeping the
emissivity-weighted temperature at its observed value (which was
deduced by assuming an isothermal model). Since the observed
surface brightness profile is still fitted by a $\beta$-model, but
now with the value $0.705 \pm 0.046$ identified as $\beta_{iso}$
(corresponding to the isothermal case, $\gamma=1$), it follows
that the $\gamma$-dependent value of $\beta$ is
$$\beta(\gamma)=4\beta_{iso}/(3+\gamma).$$ We have considered
values of $\gamma$ in the range $1$ - $5/3$. Repeating the above
procedure for the determination of $H_{0}$ for values of $\gamma$
in this range, we find that $H_{0}$ changes by at most 5\% with
respect to the value computed with $\gamma=1$. This variation is
well within the observational error quoted above, and the overall
estimated level of known systematic uncertainties stemming from
additional simplification in the modeling of the gas, such as
spherical symmetry, unclumped gas distribution, negligible impact
of CMB anisotropy, and the kinematic SZ component, as well as
other confusing signals. The additional level of systematic
uncertainties in the value of $H_0$ (deduced from SZ-X
measurements) is estimated to be $\sim 30\%$ (\eg Rephaeli 1995a,
Holzapfel {\it et al.} 1997, Birkinshaw 1999).

\section{Discussion}

Our deduced value for $H_0$ is about $\sim 30\%$ higher than the
current mean value from 33 individual results from the full data
set currently available \cite{carl02}. Given the (relatively)
large error, it is fully consistent with the recent value derived
by Spergel {\it et al.} (2003) from the first year {\it WMAP}
measurements. Clearly, the interest in using the SZ-X method to
determine $H_0$ is not diminished by the high quality {\it WMAP}
result. In addition to the need for alternative independent
methods to CMB sky maps, measurement of this basic parameter at
many redshifts and directions on the sky yields important
additional information on its variation over cosmological time,
and test of its predicted isotropy.

The full potential of the SZ-X method to determine $H_0$ has not
yet been realized. As has often been stated by Rephaeli (\eg
Rephaeli 1999), results from this method will be optimized when a
large sample of {\it nearby} clusters will be measured by
multi-frequency bolometric arrays, and with the availability of
high quality X-ray results from the {\it XMM} and {\it Chandra}
satellites. Many ground-based -- and stratospheric (such as OLIMPO
; Masi {\it et al.} 2003) -- SZ projects will collect sensitive
data on a large number of clusters. The overall impact of these
projects is likely to be larger than a single -- even if multiyear
-- satellite project: the main advantages offered by space
observations are the very low atmospheric background and a more
precise calibration based on the modulation of the CMB dipole
(when compared to calibration by planetary emission). On the other
hand, the advantages of ground-based observations (from dry sites)
are longer observation times, the use of much larger telescopes
(capable of attaining higher angular resolution), and overall
better control of systematics that can be obtained with dedicated
telescopes. Thus, even though future SZ all-sky surveys from space
experiments will sample a large quantity of clusters, best results
on individual clusters are expected from ground-based telescopes.
When the sensitivity of space observations will reach $S/N \sim
100$, it will be possible to calibrate ground-based observations
with the same precision as with space instruments. The uncertainty
in determining $H_0$ will then be comparable to that reached by
{\it WMAP} and optical measurements with the { \it Hubble Space
Telescope } (key project).

\begin{acknowledgements}

We wish to thank Naoki Itoh for the code of the analytic fitting
formula for the Gaunt factor. This work has been supported at Rome
University by COFIN-MIUR 1998, \& 2000, by ASI contract BAR, and
by a NATO Grant. Work at Tel Aviv University has been supported by
the Israel Science Foundation.

\end{acknowledgements}

\clearpage

\clearpage

\begin{deluxetable}{cccccc}
\footnotesize \tablewidth{0pt} \tablecaption{SZE experiment
parameters \label{tab:parameters}} \tablecolumns{6} \tablehead{
\colhead{Channel} & \colhead{$\nu$} & \colhead{$\Delta \nu$} &
\colhead{f.o.v. (FWHM)} & \colhead{Sensitivity} & \colhead{$\Delta
T$} \\ \colhead{} & \colhead{($GHz$)} & \colhead{($GHz$)} &
\colhead{($arcmin$)} & \colhead{($mK s^{1/2}$)} & \colhead{($\mu
K$)} }

\startdata OVRO & 32.0 & 6.5 &  7.35 & 1.40 & -520$\pm$83 \\ {\it
WMAP}(V) & 60.8 & 13.0 &  20 & 1.13 & -240$\pm$180 \\ {\it
WMAP}(W) & 93.5 & 19 & 13 & 1.48 & -340$\pm$180 \\ MITO(1) & 143 &
30 & 16 & 1.21 & -184$\pm$39 \\ MITO(2) & 214 & 30 & 16 & 1.14 &
-32$\pm$79
\\ MITO(3) & 272 & 32 & 16 & 0.896 & 172$\pm$36
\\
\enddata \label{tab:data}
\end{deluxetable}

%\begin{deluxetable}{ccccc}
%\footnotesize \tablewidth{0pt} \tablecaption{$y$ and $H_{0}$
%\label{tab:ips}} \tablecolumns{5} \tablehead{ \colhead{$\gamma$} &
%\colhead{$y\cdot10^{5}$} & \colhead{$\Delta y\cdot10^{5}$} &
%\colhead{$H_{0}$} & \colhead{$\Delta H_{0}$}}
%
%\startdata 1.00 & 8.60 & 1.10 &84.1 & 25.7
%\\ 1.05 & 8.57 & 1.16 & 83.9 & 26.0
%\\ 1.10 & 8.57 & 1.15 & 83.2 & 25.7
%\\ 1.15 & 8.57 & 1.14 & 82.8 & 25.7
%\\ 1.20 & 8.57 & 1.14 & 82.6 & 25.7
%\\ 1.25 & 8.57 & 1.13 & 82.5 & 25.8
%\\ 1.30 & 8.58 & 1.12 & 82.6 & 25.8
%\\ 1.35 & 8.59 & 1.12 & 82.8 & 26.0
%\\ 1.40 & 8.60 & 1.12 & 83.0 & 26.1
%\\ 1.45 & 8.61 & 1.12 & 83.3 & 26.3
%\\ 1.50 & 8.62 & 1.11 & 83.7 & 26.4
%\\ 1.55 & 8.63 & 1.11 & 84.2 & 26.7
%\\ 1.60 & 8.64 & 1.11 & 84.7 & 27.0
%\\ 1.65 & 8.65 & 1.11 & 85.3 & 27.3\\
%\enddata \label{tab:H} \end{deluxetable}


\begin{thebibliography}

\bibitem[Battistelli {\it et al.} 2002]{batt02}
{Battistelli}, E.S., {De Petris}, M., {Lamagna}, L., {Melchiorri},
F., {Palladino}, E., {Savini}, G., {Cooray}, A., {Melchiorri}, A.,
{Rephaeli}, Y. and {Shimon}, M., 2002, {\apj}, 580, L101-L104

\bibitem[Bennett {\it et al.} 2003a]{ben03a}
{Bennett}, C.L., {Halpern}, M., {Hinshaw}, G., {Jarosik}, N.,
{Kogut}, A., {Limon}, M., {Meyer}, S.S., {Page}, L., {Spergel},
D.N., {Spergel}, D.N., {Tucker}, G.S., {Wollack}, E.,  {Wright},
E.L., {Barnes}, C., {Greason}, M.R., {Hill}, R.S., {Komatsu}, E.,
{Nolta}, M.R., {Odegard}, N., {Peirs}, H.V., {Verde}, L., and ,
{Weiland}, L., 2003a, {\apjs}, 148, 1, 1-27

\bibitem[Bennett {\it et al.} 2003b]{ben03b}
{Bennett}, C., {Hill}, R.S., {Hinshaw}, G., {Nolta}, M.R.,
{Odegard}, N., {Page}, L., {Spergel}, D.N., {Weiland}, J.L.,
{Wright}, E.L., {Halpern}, M., {Jarosik}, N., {Kogut}, A.,
{Limon}, M., {Meyer}, S.S., {Tucker}, G.S., and {Wollack}, E.,
2003b, {\apjs}, 148, 1, 97-117

\bibitem[Birkinshaw 1999]{birk99}
{Birkinshaw}, M., 1999, {\physrep}, 310, 97

\bibitem[Birkinshaw {\it et al.} 1981]{birk81}
{Birkinshaw}, M., {Gull}, S.F., {Northover}, K.J.E., 1981,
{MNRAS}, 197, 571

\bibitem[Carlstrom {\it et al.} 2002]{carl02}
{Carlstrom}, J.R., {Holder}, G.P., and {Reese}, E.D., 2002,
{\araa}, 40, 643

\bibitem[Cavaliere \& Fusco-Femiano 1976]{Cava76}
{Cavaliere}, A., \& {Fusco-Femiano}, R., 1976, {A\&A}, 49, 137

\bibitem[Colafrancesco {\it et al.} 2003]{cola03}
{Colafrancesco}, S., {Marchegiani}, P., and {Palladino}, E., 2003,
{A\&A}, 397, 27

\bibitem[DePetris {\it et al.} 2002]{depe02}
{De Petris}, M., {D'Alba}, L., {Lamagna}, L., {Melchiorri}, F.,
{Orlando}, A., {Palladino}, E., {Rephaeli}, Y., {Colafrancesco},
S., {Kreysa}, E., and {Signore}, M., 2002, {\apj}, 574, L119-L122

\bibitem[Furuzawa {\it et al.} 1998]{furu98}
{Furuzawa}, A., {Tawara}, Y., {Kunieda}, H., {Yamashita}, K.,
{Sonobe}, T., {Tanaka}, Y., and {Mushotzky}, R., 1998, {\apj},
504, 35-41

\bibitem[Herbig {\it et al.} 1995]{her95}
{Herbig}, T., {Lawrence}, C.R., and {Readhead}, A.C.S., 1995,
{\apjl}, 449, L5

\bibitem[Holzapfel {\it et al.} 1997]{Holz97}
{Holzapfel}, W.L., {Arnaud}, M., {Ade}, P.A.R., {Church}, S.E.,
{Fischer}, M.L., {Mauskopf}, P.D., {Rephaeli}, Y., {Wilbanks},
T.M., and {Lange}, A.E., 1997, {\apj}, 480, 449

\bibitem[Hughes \& Birkinshaw 1998]{Hugh98}
{Hughes}, J.P., \& {Birkinshaw}, M., 1998, {\apj}, 501, 1

\bibitem[Itoh {\it et al.} 1998]{itoh98}
{Itoh}, N., {Kohyama}, Y., and {Nozawa}, S., 1998, {\apj}, 502,
7-15

\bibitem[Itoh {\it et al.} 2002]{Itoh02}
{Itoh}, N., {Sakamoto}, T., {Kusano}, Y., {Kawana}, Y., and
{Nozawa}, S., 2002, {A\&A}, 382, 722-729

\bibitem[Itoh {\it et al.} 2000]{itoh00}
{Itoh}, N., {Sakamoto}, T., {Kusano}, Y., {Nozawa}, S.,and
{Kohyama}, Y., 2000, {\apjs}, 128, 125-138

\bibitem[Masi {\it et al.} 2003]{masi2003}
{Masi}, S., {Ade}, P.A.R., {Boscaleri}, A., {de Bernardis}, P.,
{De Petris}, M., {De Troia}, G., {Fabrini}, M., {Iacoangeli}, A.,
{Lamagna}, L., {Lange}, A.E., {Lubin}, P., {Mauskopf}, P.D.,
{Melchiorri}, A., {Melchiorri}, F., {Nati}, F., {Nati}, L.,
{Orlando}, A., {Piacentini}, F., {Pierre}, F., {Pisano}, G.,
{Polenta}, G., {Rephaeli}, Y., {Romeo}, G., {Salvaterra}, L.,
{Savini}, G., {Valiante}, E., and {Yvon}, D., 2003, to appear in
the proceedings of the 4th National Conference on Infrared
Astronomy, 4-7 December 2001, Perugia, eds. S. Ciprini, M. Busso,
G. Tosti and P. Persi, "Memorie della Societa' Astronomica
Italiana", 74, 1

\bibitem[Mason {\it et al.} 2001]{mason01}
{Mason}, B.S., {Myers}, S.T., and {Readhead}, A.C.S., 2001,
{\apjl}, 555, L11

\bibitem[Mohr {\it et al.} 1999]{Mohr99}
{Mohr}, J.J., {Mathiesen}, B., and {Evrard}, A.E., 1999, {\apj},
517, 2, 627

\bibitem[Nozawa, Itoh, \& Kohyama 1998]{Noza98}
{Nozawa}, S., {Itoh}, N., \& {Kohyama}, Y., 1998, {\apj}, 508, 17

\bibitem[Reese {\it et al.} 2002]{reese02}
{Reese}, E.D., {Carlstrom}, J.E., {Joy}, M., {Mohr}, J.J.,
{Grego}, L., and {Holzapfel}, W.L., 2002, {\apj}, 581, 53

\bibitem[Reese {\it et al.} 2000]{reese00}
{Reese}, E.D., {Mohr}, J.J., {Carlstrom}, J.E., {Joy}, M.,
{Grego}, L., {Holder}, G.P., {Holzapfel}, W.L., {Hughes}, J.P.,
{Patel}, S.K., and {Donahue}, M., 2000, {\apj}, 533, 38

\bibitem[Rephaeli 1995a]{reph95a}
{Rephaeli}, Y., 1995a, {\araa}, 33, 541

\bibitem[Rephaeli 1995b]{reph95b}
{Rephaeli}, Y., 1995b, {\apj}, 445, 33

\bibitem[Rephaeli 1999]{reph99}
{Rephaeli}, Y., 1999, {`3K cosmology', L. Maiani, F. Melchiorri,
\& N. Vittorio, eds., AIP}, 476, 310

\bibitem[Savini {\it et al.} 2003]{sav03}
 {Savini}, G., {Orlando}, A., {Battistelli}, E.S., {De Petris}, M.,
{Lamagna}, L., {Luzzi}, G., and {Palladino}, E., 2003, {New Astronomy},
8, 7, 727-736

\bibitem[Shimon \& Rephaeli 2003]{shim03} {Shimon}, M. \& {Rephaeli}, Y.,
2003, {New Astronomy},  in press, (astro-ph/0309098)

\bibitem[Silverberg {\it et al.} 1997]{sil97}
{Silverberg}, R. {Cheng}, E.S., {Cottingham}, D.A., {Fixsen},
D.J., {Inman}, C.A., {Kowitt}, M.S., {Meyer}, S.S., {Page}, L.A.,
{Puchalla}, J.L., {Rephaeli}, Y.,  1997, {\apj}, 485, 22

\bibitem[Spergel {\it et al.} 2003]{spe03}
{Spergel}, D.N., {Verde}, L., {Peiris}, H.V., {Komatsu}, E.,
{Nolta}, M.R., {Bennett}, C.L., {Halpern}, M., {Hinshaw}, G.,
{Jarosik}, N., {Kogut}, A., {Limon}, M., {Meyer}, S.S., {Page},
L., {Tucker}, G.S., {Weiland}, J.L., {Wollack}, E., and {Wright},
E.L. 2003, {\apjs}, 148, 1, 175-194

\bibitem[Sunyaev \& Zel'dovich 1972]{SunZel72}
{Sunyaev}, R.A. \& {Zel'dovich}, Ya. B., 1972,  {Comm. Astrphys.
Space Phys.}, 4, 173

\end{thebibliography}
\end{document}